# Power scaling of high efficiency 1.5micron cascaded Raman fiber lasers

V. R. Supradeepa,[1,*] and Jeffrey W. Nicholson[1]

[1]OFS Laboratories, 19 Schoolhouse Road, Suite 105, Somerset, New Jersey 08873, USA
*Corresponding author: supradeepa@ofsoptics.com

High power fiber lasers operating at the 1.5micron wavelength region have attractive features like eye-safety and atmospheric transparency, and cascaded Raman fiber lasers offer a convenient method to obtain high power sources at these wavelengths. A limitation to power scaling however has been the lower conversion efficiency of these lasers. We recently introduced a high efficiency architecture for high power cascaded Raman fiber lasers applicable for 1.5micron fiber lasers. Here we demonstrate further power scaling using this new architecture. Using numerical simulations we identify the ideal operating conditions for the new architecture. We demonstrate a high efficiency 1480nm cascaded Raman fiber laser with an output power of 301 W, comparable to record power levels achieved with rare-earth doped fiber lasers in the 1.5 micron wavelength region.

High power, single mode fiber lasers at the 1.5 micron wavelength region have the attractive features of eye safety and enhanced atmospheric transmission. This makes them interesting for a variety of commercial, government, aerospace and defense applications. ErYb co-doped fibers pumped at 975nm (where mature, low-cost diode technology is available) has been the standard method to generate high powers in the 1.5 micron wavelength. Owing to the large quantum defect between the pump and signal wavelengths, the thermal load is significantly enhanced which becomes a limitation for power scaling. More importantly, at higher powers, parasitic lasing by the Yb ions significantly reduces efficiency and limits power scaling. The highest power, demonstrated, in 2007, was 297W, obtained for a total pump power of 1.2kW [1]. The slope efficiency was < 20% at maximum power due to parasitic Yb lasing. This limitation to scalability has prevented any improvement since then. Other techniques involving Yb free Er doped fibers are limited in power scaling or output power due to enhanced thermal load (if pumped at 980nm) [2] or lack of cost-effective and efficient pump sources (for pumping in-band) [3].

Cascaded Raman fiber lasers provide a convenient alternative to generate high powers at 1.5 micron as well as other wavelengths [4]. Furthermore, Raman lasers can also serve as high brightness, low quantum defect pump sources for Erbium-doped fiber lasers and amplifiers [5-8]. In-band core pumping reduces the nonlinearity significantly by requiring much shorter length of gain media and is particularly attractive for single frequency and short pulse Er-doped fiber amplifiers. The principle behind Raman fiber lasers is wavelength conversion of the output of a rare-earth doped fiber laser (in this case the mature technology of Yb-doped fiber lasers) to the required output wavelength using a series of Raman Stokes shifts. Conventionally, the wavelength conversion is achieved using a cascaded Raman resonator - a series of nested cavities at each of the intermediate Stokes wavelengths comprising high reflectivity fiber Bragg gratings and a high nonlinearity Raman fiber. Conversion is terminated with a low reflectivity output coupler at the final wavelength. At the output most of the light is at the final wavelength with small residual fractions at all the intermediate wavelengths.

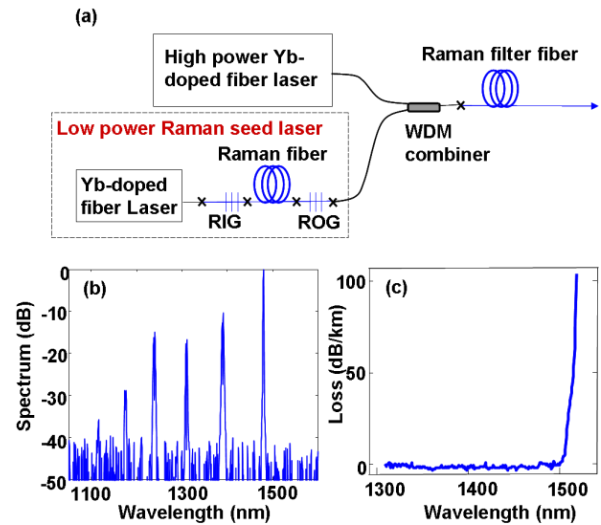

Fig. 1. (a) Experimental scheme, RIG/ROG – Raman input and output grating sets, WDM – wavelength division multiplexer (b) Representative spectrum of the seed source (c) Transmission characteristic of the Raman filter fiber

Power scaling of Raman fiber lasers is a subject of current interest [9, 10]. Using the conventional architecture [4] over 100W at 1480 nm was achieved, starting from a Yb-doped fiber laser at 1117nm [11]. A Raman filter fiber based on a W-shaped index profile with high loss above 1500 nm (see Fig. 1c) aided in the power scaling by suppressing additional Stokes scattering to wavelengths longer than 1480 nm [10]. However the optical–to-optical conversion efficiency from Yb fiber laser output (1117nm) to 1480 nm Raman output was only ~48%, far below the quantum efficiency limit of 75%. Excess losses associated with the cascaded Raman resonator assembly were identified to be responsible for the reduced efficiency.

Recently, a new, high efficiency architecture was demonstrated based on a single pass cascaded Raman amplifier, seeded at all intermediate Stokes wavelengths [12]. Fig 1(a) shows the architecture. A seed source comprising a low power conventional cascaded Raman laser is power combined with a high power Yb-doped fiber laser and sent through the Raman fiber. The low power cascaded Raman laser simultaneously provides all the necessary wavelengths due to the presence of residual intermediate Stokes orders. Fig 1(b) shows the output spectrum with only the seed source turned on. Using this architecture, 204W at 1480nm was demonstrated with a record 1117nm to 1480nm conversion efficiency of 65% (~43% from 975nm to 1480nm, including the conversion efficiency of the Yb-doped fiber laser).

In this paper, we explore further the ideal operating conditions and power scaling aspects of this architecture. Experimentally, we demonstrate a record output power of 301W at 1480nm while maintaining high conversion efficiency limited only by available input power. Furthermore, this laser has significantly higher conversion efficiencies compared to other 1.5micron fiber laser sources at similar power levels.

A simple way to scale the output power (sacrificing the efficiency to some extent) is to increase the output power in the seed source. In this section, we explore using simulations, increasing the seed power and identifying proper operating conditions for the architecture. Let $P_i, i = 0,1.....n$ be the power components in each of the wavelength components, $P_0$ is the pump, $P_{n-1}$ is the signal component and $P_n$ is the undesirable wavelength created by further conversion of the signal. For the 1117nm to 1480nm conversion in this work, $n = 6$. Since the conversion is seeded at all intermediate wavelengths, we can ignore the amplified spontaneously generated backward components. The equations governing each Stokes components are [13]

$$\frac{dP_i}{dz} = -\alpha_i P_i + \frac{g_{i-1}}{A_{eff_{i-1}}} P_i P_{i-1} - (\frac{\lambda_i}{\lambda_{i-1}}) \frac{g_i}{A_{eff_i}} P_i P_{i+1} \quad (1)$$

where $\alpha_i$ is the loss for each wavelength, $A_{eff_i}$ is the effective area of the fiber at each of the Stokes wavelengths, $g_i$ is the Raman gain coefficient and the term $\lambda_i / \lambda_{i-1}$ represents the quantum defect. The above equation has 3 terms corresponding to linear loss, growth due to gain from previous Stokes wavelength and loss due to further conversion to the next Stokes wavelength. For the pump $P_0$, there is no gain component and the equation is modified to

$$\frac{dP_0}{dz} = -\alpha_0 P_0 - (\frac{\lambda_1}{\lambda_0}) \frac{g_0}{A_{eff_0}} P_0 P_1 \quad (2)$$

And for the final Stokes component created by further conversion of the signal wavelength, we can ignore the term governing conversion to the next Stokes order and the equation is modified to

$$\frac{dP_n}{dz} = -\alpha_n P_n + \frac{g_{n-1}}{A_{eff_{n-1}}} P_{n-1} P_n \quad (3)$$

In our work, $A_{eff}$ increases from $\sim 12 \mu m^2$ at 1117nm to $\sim 18 \mu m^2$ at 1480nm. All the wavelength components are unpolarized and for the Raman gain coefficient correspondingly, [13]

$$g_i = 0.5 \times 10^{-13} \times \frac{1 \mu m}{\lambda_i} \quad (4)$$

Where $\lambda_n$ is in microns. Linear loss $\alpha_i, i = 0,1,.....,(n-1)$ is assumed to be 1dB/km. $\alpha_n$, the excess loss for the converted signal component introduced by the Raman filter fiber needs to be high enough to suppress further conversion. From eqn (3) we see that if

$$\alpha_n > (g_{n-1} / A_{eff_{n-1}}) P_{n-1} \quad (5)$$

the spurious component (at 1590nm) cannot grow. For an output power of 300W at 1480nm, this corresponds to ~3dB/m. The filter fiber we use provides a higher loss value than this at 1590nm suppressing all further conversion of the signal light.

The pump wavelength is provided by the high power Yb-doped fiber laser and intermediate and final wavelengths are provided by the cascaded Raman seed laser. The high power Yb-doped fiber laser provides up to 450W of output power at 1117nm, and the intermediate Stokes components in the seed source is 15-dB below the power at its final wavelength, corresponding to experimental conditions. The simulation results were found to be insensitive to the exact power levels in each of the intermediate Stokes components. Since all the components in the above equations are forward moving, they can be easily solved starting from the initial values by evaluating the derivative at each length step and updating the components for the successive step on a discretized length grid. Of interest is the variation of conversion efficiency and total output power as the power in the seed source is varied.

Fig 2(a) shows the output power at 1480nm as the power in the seed source is varied from 0.1W to 100W for three appropriately chosen lengths of the Raman filter fiber. Also shown is the optimal conversion in the amplifier given by quantum limited conversion together with the power in the seed source. As expected, when the seed source power is too low there is incomplete conversion. This can be compensated to some extent by using longer fiber lengths with the addition of a small increase to linear loss in the fiber. There is interesting behavior observed as the seed source power is increased. The output power shows saturation like behavior. This progressively increases the difference between the optimal conversion and obtained output power, indicating that power scaling cannot be achieved by simply increasing the power in the seed source. Fig 2(b) shows the corresponding plots in terms of conversion efficiency in the amplifier as a function of seed source power. The seed source power beyond the 10W level results in progressively reduced conversion efficiencies and in the 1-10W level, close to quantum limited efficiency can be achieved. Thus, having a moderate power seed source in the range of 1 to 10 W is optimal for achieving high conversion efficiency. In these simulations we have ignored excess losses caused by the combiner component and the splice to small effective area Raman fiber which would be present in the actual system.

from 10W to 40W obtained with a slightly longer Raman fiber length of 50m. The incomplete power transfer out of 1310nm is clearly seen as the seed source power is increased. The 1310 component relative to the 1480nm component (zoomed in the inset) nearly triples in height from a ratio of 0.05 to 0.15.

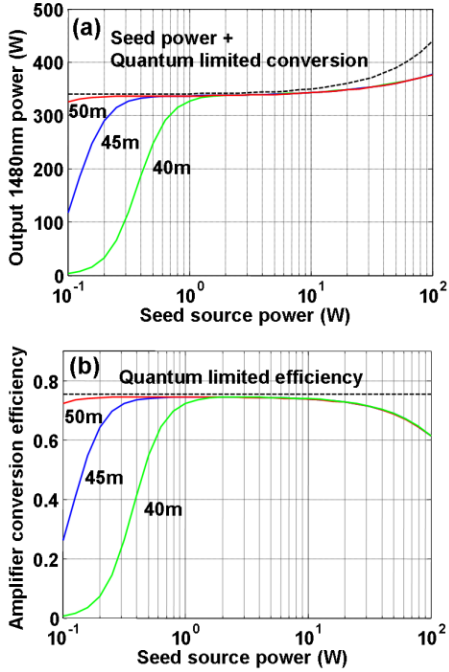

Fig. 2. (a) Output power as a function of total power of the seed source for 3 different lengths. Also shown is the optimal power output of the amplifier (b) Conversion efficiency in the amplifier stage as a function of seed source power

We can understand the origins of this behavior by looking at the growth of individual Stokes components as a function of input power for various seed source powers. Fig 3(a) shows the case for a 45m long cascaded Raman amplifier with a seed source power of 10W which from Fig 2 achieves a high degree of conversion efficiency. There is a progressive growth and decay of all the intermediate Stokes components except the 1390nm component which is slightly suppressed. At maximum power, a high degree of wavelength conversion with most of the power in the final wavelength is achieved. Fig 3(b) shows the situation when the seed source power is increased to 40W. The growth of individual components is shifted to lower power owing to higher seed powers. The penultimate component however, at 1390nm is further suppressed, due to the presence of significant power at 1480nm from the start. The consequence is incomplete power transfer out of 1310nm resulting in incomplete wavelength conversion and hence reduced efficiency. Fig 3(c) shows experimentally measured spectra at an output power of ~200W at 1480nm (corresponding to an input power of ~300W at 1117nm) as the seed source power is increased

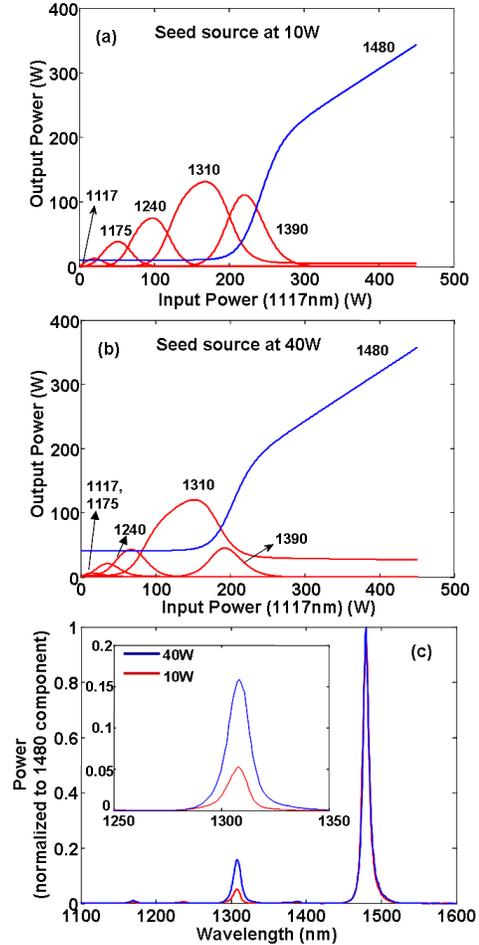

Fig.3. (a), (b) The evolution of power at all the Stokes wavelength components as a function of input power for different seed source powers and Raman fiber length of 45m. (c) Experimentally measured spectra at an output power of ~200W at 1480nm as the seed source power is increased from 10W to 40W for a fiber length of 50m

From the simulations and experiments above, to achieve power scaling, increasing the power in the Yb-doped fiber laser while maintaining the seed laser at low powers is optimal for conversion efficiency. This configuration has the added advantage that a low power seed laser enables the total conversion efficiency to be close to that of the cascaded Raman amplifier.

In the experiment, the high power Yb-doped fiber laser provided a maximum of 453 W at 1117nm. It was based on a forward pumped oscillator and a two stage bi-directionally pumped power amplifier configuration. Standard 6+1 to 1 pump combiners and 6/125micron Yb-doped gain fiber from OFS [14] were used and the diode modules were nominally 25W at 975nm. The conversion efficiency of the Yb-doped fiber laser (975nm pump diodes to 1117nm) was ~66%. We used a 10W cascaded Raman laser as the seed source and 45m of Raman filter fiber. From the simulations above, this should achieve close to quantum limited conversion efficiency. There were two additional sources of loss in the system. The WDM combiner had ~4% loss owing to its isolation properties and the splice from standard single mode fiber to the low effective area Raman fiber had an additional 5% loss.

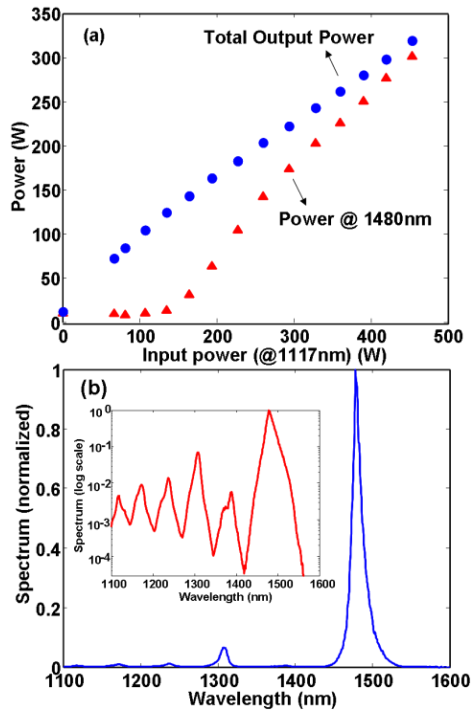

Fig. 4. (a) Total output power and 1480nm component as a function of input power at 1117nm (b) Output spectrum in linear and log scale at maximum power.

Fig 4(a) shows the total output power and the 1480 nm component as a function of input power at 1117 nm. We achieve an output power of ~301 W at 1480 nm (limited only by input power) for a total input power of ~470 W (including the Yb-doped fiber laser in the seed source) with a total conversion efficiency of ~64% (1117nm to 1480nm, quantum limited efficiency of 75%). Taking into account the efficiency of the Yb-doped fiber laser, we achieve an optical to optical conversion efficiency from 975 nm pump to 1480 nm signal of ~42%. In comparison, the conversion efficiency in [1] was < 25% and decreasing due to parasitic lasing. Yb-doped fiber lasers with over 80% efficiency [15] have been demonstrated and there is significant room to improve this in our system. We believe incorporating this, the net efficiency (from 975nm to 1480nm) can be improved to >50%. Fig 4(b) shows the measured output spectrum at full power (log scale in the inset). More than 95% of the power is in the 1480 nm band indicating a high level of wavelength conversion while high suppression of the next Stokes order at 1590 nm is maintained through the use of the filter fiber. The reduction in efficiency below the quantum limit can be accounted for by the loss of the WDM combiner, splices and residual power in the other Stokes components, giving clear directions to pursue further improvements in performance.

In summary, we have investigated power scaling and efficiency aspects of the new cascaded Raman laser architecture as a function of seed source power. An interesting behavior of power saturation and reduced efficiency was observed with increased seed source power and the physical basis behind this was identified. Using close to optimal conditions as obtained from the simulations, we demonstrated a 1480nm cascaded Raman laser with a pump limited output power of 301W and conversion efficiencies of 64% in the Raman stage from 1117nm to 1480 and 42% overall (from 975nm to 1480nm). We believe this is the highest power Raman fiber laser at the 1.5micron wavelength, achieving an output power comparable to record ErYb results, but with much higher optical to optical efficiency.

We would like to thank Dr. Kazi Abedin for helpful discussions.